%% file: hyperloop_system_optimization.tex
\documentclass[journal]{aiaa-pretty}

\usepackage{amsmath, amssymb}
\usepackage{standalone}
\usepackage{tikz}
\usetikzlibrary{decorations.pathreplacing}

\author[Kirschen and Burnell]{%
	Philippe G. Kirschen\thanks{Manager of System Optimization, \texttt{philippe@virginhyperloop.com}},
	Edward Burnell\thanks{Partner, Convex Technologies, \texttt{partners@convex.tech}}\\
	\textit{Virgin Hyperloop, Los Angeles, CA 90021}}

\title{Hyperloop System Optimization}

\abstract{%
Hyperloop system design is a uniquely coupled problem because it involves the
simultaneous design of a complex, high-performance vehicle and its accompanying
infrastructure.
In the clean-sheet design of this new mode of high-speed mass transportation
there is an excellent opportunity for the application of rigorous system
optimization techniques.
This work presents a system optimization tool, HOPS, that has been
adopted as a central component of the Virgin Hyperloop design process.
We discuss the choice of objective function, the use of a convex
optimization technique called geometric programming, and the level of
modeling fidelity that has allowed us to capture the system's many intertwined,
and often recursive, design relationships.
We also highlight the ways in which the tool has been used.
Because organizational confidence in a model is as vital as its technical
merit, we close with discussion of the measures taken to build stakeholder
trust in HOPS.
}

\begin{document}
\maketitle

\section{Introduction}
\label{introduction}

The hyperloop is a concept for a high-speed mass transportation system that
uses an enclosed low-pressure environment and small autonomous vehicles
(``pods'') to enable an unparalleled combination of short travel times, low
energy consumption, and ultra-high throughput.
From an engineering perspective, hyperloop system design is a highly-coupled
optimization problem with many recursive design relationships.
It is also a clean-sheet design problem -- there is no operational precedent to
use as a baseline -- so there is no ``initial guess'' from which to begin.
It is ripe for the application of rigorous system optimization.

Hyperloop system designers must answer questions such as:
How fast should the pod travel?
How quickly should the pod accelerate?
How many passengers should each pod be able to carry?
How heavy does a pod need to be to achieve these things?
How big should the tube be?
How low should the air pressure inside the tube be?
How big does each portal need to be?
How big does a pod fleet need to be?
The answers to these questions and many more are not only important to enable
detailed design but they are also inextricably intertwined.

To further complicate matters, the goal is to design a superlative mode of
transportation; one that offers the shortest journey time, highest
departure frequency, lowest energy consumption, cheapest ticket price,
highest passenger throughput, and highest levels of safety and comfort.
We therefore need a disciplined way of trading between these often-competing
objectives and making engineering decisions based on their impact on an
appropriately chosen objective function.

Fortunately, advances in powerful mathematical solvers and
user-friendly optimization modeling languages have made it possible to
formulate, modify, and solve large and complex optimization problems quickly
and in a way that is approachable and useful in a practical engineering
context.

We have leveraged such advances to create a hyperloop system optimization tool
that has been used extensively in the development of the Virgin Hyperloop
system.
The purpose of this paper is to describe both the hyperloop system optimization
problem in general, and the tool we have developed.

The structure of the paper is as follows.
First, to provide context, we give a brief overview of the defining
characteristics of a hyperloop system.
Next, we extend this to explain what makes the design of such a system not only
an interesting optimization problem but also an extremely tightly coupled
one that is uniquely well suited to formal and rigorous system
optimization.
We then explain how this motivated the choice of a particular optimization
technique.
The remainder of the paper describes HOPS\footnote{Regrettably, this acronym
expands to Hyperloop OPtimization Software.}, a hyperloop system optimization
tool that allows engineers to answer the questions above (and many more)
using convex optimization.
This includes a discussion of the choice of objective function, a high-level
description of the major subsystem models, examples of how the tool has been
used, and finally a description of how the tool was implemented in, and adopted
by, the company.

\section{Hyperloop}

Although Elon Musk coined the name and
repopularized\cite{goddard1950vacuum,oster2011evacuated,bierlaire2001acceptance}
the concept with the Hyperloop Alpha white paper\cite{hyperloopalpha}, the
notion of what constitutes a hyperloop has evolved significantly since its
publication in 2013, with different companies pursuing different system
architectures.
Time will tell which architecture(s), if any, will be commercially successful;
however, all of the most promising architectures share certain key features.

By virtue of these features, we claim that hyperloop system design presents a
singular opportunity for the application of multidisciplinary design
optimization (MDO), agnostic of the choice of architecture.
However, we also claim that using an MDO tool that considers the full
system-wide impacts of each design choice is crucial to choosing the best
system architecture.

\subsection{The Key Features of a Hyperloop System}

To achieve the high-level objectives of low travel time, low wait time, low
energy consumption and high throughput, a hyperloop system must have the
following key features:
\begin{enumerate}
	\item Small\footnote{How small (i.e. how many passengers) is obviously
		one of the most important variables to optimize, but
		intuitively we can say that they should carry more people than
		a car and fewer people than a regional jet.} pods to allow
		highly demand-responsive direct-to-destination service.
	\item An enclosed low pressure environment to enable low energy
		consumption despite high speeds and small pods.
	\item ``Pod-side switching'', i.e. no moving parts on the wayside, to
		allow ultra-high throughputs\footnote{Ultra-high throughput is
		not only important for designing a system that is equipped to
		handle the demands of future population growth, but it is also
		a key part of reducing the total cost per passenger by enabling
		much higher utilization than a conventional rail or maglev
		system.} despite small pods and direct-to-destination service.
\end{enumerate}

\begin{figure}[ht]
        \centering
	\includegraphics[width=\linewidth]{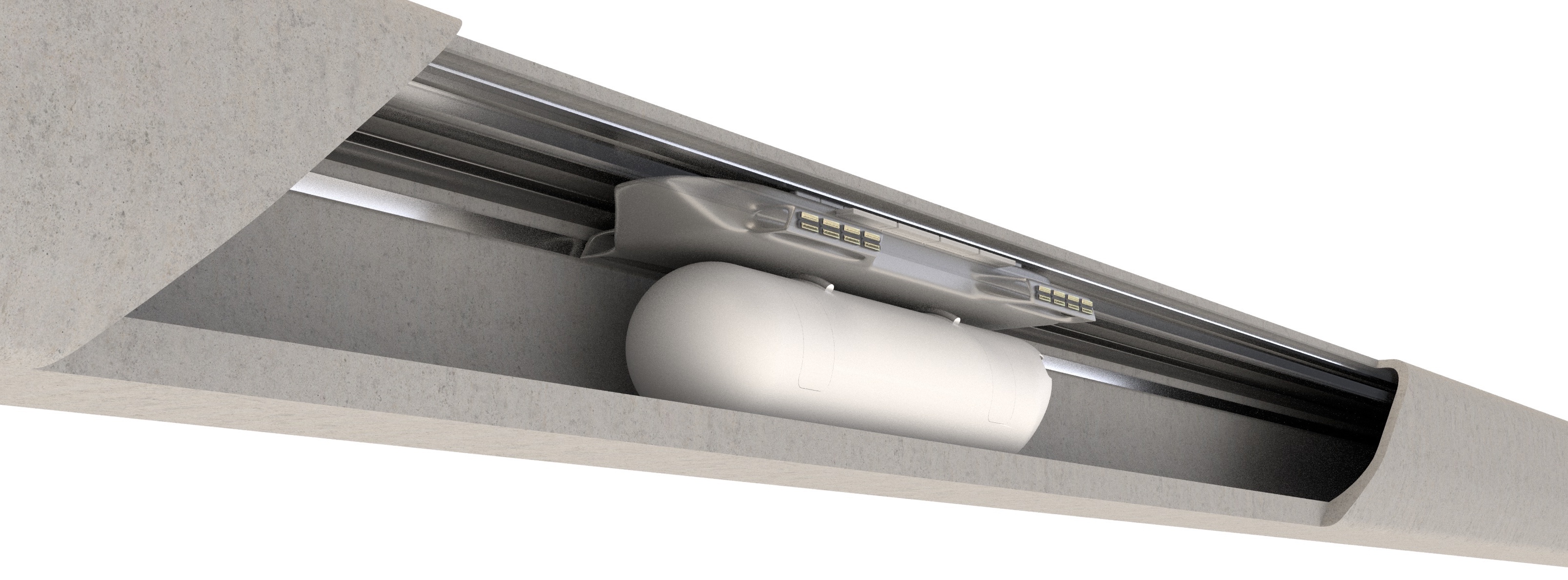}
	\caption{A hyperloop pod in a tube cutaway. The pod comprises a
	fuselage and a bogie. The tube encloses the low pressure environment and
	supports track elements.}
	\label{fig:system-diagram}
\end{figure}

There are four major elements of a hyperloop system: the pod fleet, the
linear infrastructure (``hyperstructure''), the pressure
management system, and the stations (``portals'').
The pod fleet carries passengers to their destinations.
The hyperstructure encloses the low pressure
environment and supports any necessary track elements for propulsion,
levitation, guidance, and emergency braking.
The enclosed nature of the system also allows it to be more resilient to
weather and safety hazards than other transportation systems.
The pressure management system establishes and maintains a low-pressure
environment inside the tube.
The portals provide an interface between the hyperloop system and the outside
world; they are where passengers board and disembark pods and where pods'
resources are replenished.
\autoref{fig:system-diagram} shows a rendering of a pod inside a tube
to help visualize these key elements.

\subsection{A Singular Opportunity for Multidisciplinary Design\\ Optimization}

Aircraft design is often cited as the exemplar of an interesting and
challenging multidisciplinary optimization problem because of its conflicting design
pressures from different technical domains (e.g aero-structural optimization of
a wing), its recursive design relationships (e.g. the size of a wing depends on
the weight of the aircraft which depends on the weight of the wing), and its strong
coupling between subsystems (e.g. for a multi-engine aircraft the size of the
vertical tail might depend on the thrust of the
engines)~\cite{henderson2012aircraft,kroo1994multidisciplinary}.
As a result, a vast amount of research has gone into developing tools for
multidisciplinary design optimization of aircraft
~\cite{martins2013multidisciplinary,martins2004high,sobieszczanski1997multidisciplinary}.

We claim that hyperloop system design presents a singular opportunity for
multidisciplinary design optimization because of four properties: it requires
modeling from many engineering disciplines, it is tightly coupled and highly
recursive, it is a clean-sheet design problem, and the behavior of the system
is expected to be highly deterministic.

\subsubsection{Multidisciplinary}
A hyperloop system invokes many engineering disciplines including structures,
electronics, aerodynamics, electromagnetics, thermodynamics, controls,
manufacturing, and civil engineering.
Because it is a mass transportation system and involves large infrastructure
projects, there are also important economic models to capture.

\subsubsection{Coupled and Recursive}
Hyperloop system design is an even more coupled problem than aircraft design.
The most obvious reason for this is that the environment in which pods travel
and the structure containing this environment are also key parts of the design
and they affect, and are affected by, the design of the pod and the pressure
management system.
For example, the tube pressure and blockage ratio (pod-to-tube-area ratio) are
key drivers of aerodynamic drag and therefore pod energy consumption, while the
tube is one of the most significant drivers of total cost.
Similarly, the size and cost of track elements are also directly linked to the
design and performance of their counterpart pod subsystems.

Whilst airports and railway stations can impact the design of their respective
vehicles, hyperloop portals are a more tightly coupled form of passenger
terminal, because their size and cost are more closely linked to the size of
pods and the pod turnaround time.
For battery-powered pods, turnaround time is often gated by
charging rather than passenger boarding/disembarking, which introduces a
coupling between portal size and pod performance that is unique for a mass
transportation system.

Pods do not operate in isolation but rather as part of a fleet, and this adds
another source of coupling, with both favorable and adverse effects.
For example, more pods operating close together can reduce average aerodynamic
drag, but having more pods in the system also means more sources of air leak
into the tube, which can increase the optimal tube pressure.

Finally, not only do all of the major subsystems have recursive design
relationships (e.g. the propulsion system mass depends on the pod mass which
depends on the propulsion system mass) but they are also even more coupled to
each other than subsystems on vehicles that operate in atmospheric pressure.
Because there is very little convective cooling in a near-vacuum environment,
the thermal management system plays a significant role in system design by
introducing a strong coupling between efficiency and mass; even small decreases
in efficiency can cause a significant ``mass spiral'' effect under certain
conditions.

\begin{figure*}[ht]
        \centering
	\includegraphics[width=0.8\textwidth]{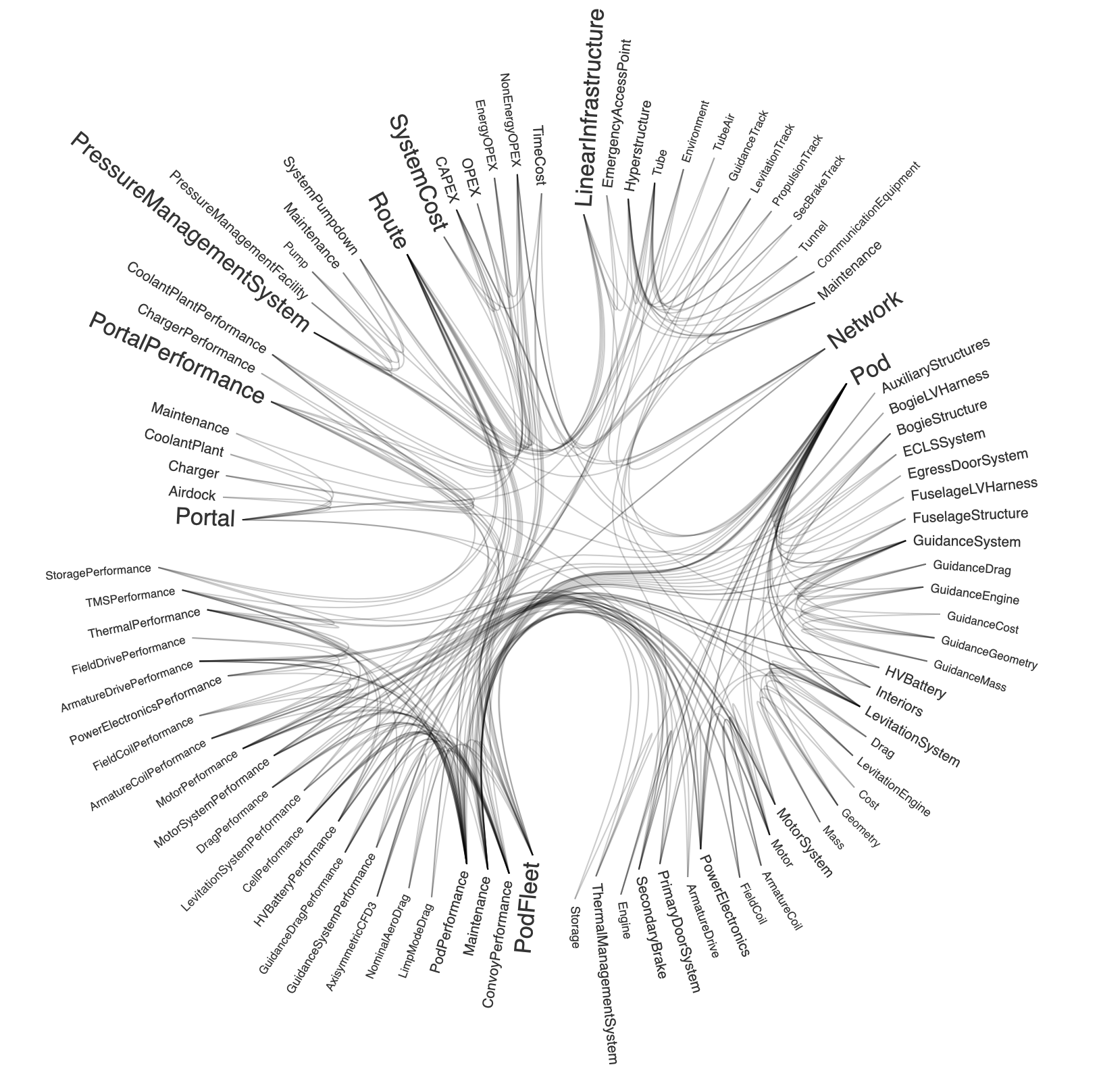}
	\caption{Model coupling in HOPS. Each labeled endpoint represents a
	model and each arc represents a coupling between two models in the
	form of shared variables.  The radial position of model names indicates
	their position in the system hierarchy, illustrated more clearly in
	\autoref{fig:model_structure}.}
	\label{fig:model_links}
\end{figure*}

\autoref{fig:model_links} captures the coupling between system elements and pod
subsystems.
Every line represents the interdependence of two models, with variables flowing
to and/or from the connected models.

It should be noted that not every potential hyperloop system architecture has
the same degree of coupling, and while more coupling can make design more
difficult, it is often a byproduct of architecture choices that unlock other
key advantages.

\subsubsection{Deterministic}
Another factor that makes hyperloop design well-suited to rigorous system
optimization is that, as an autonomous, track-based system operated in a
controlled environment, it has a well-defined nominal operation for any given
route; the environmental conditions can be well characterized and the energy
consumption and thermal profiles are therefore easy to model with relative
confidence.
In other words, operations are highly deterministic.
While there are other sources of uncertainty and some subsystems will
necessarily be sized by off-nominal scenarios, the pod design is not subject to
extreme, rare, and/or difficult-to-analyze weather phenomena and can largely be
optimized around well-defined nominal operations.

\subsubsection{Clean-Sheet Design}
The other rare optimization opportunity afforded to hyperloop design at this
time is the lack of any legacy infrastructure or standards that need to be
accommodated.
Hyperloop system design is a clean-sheet problem, governed by
physics, economics, and human psychology.
As a nascent industry, commercial success is also not (yet) dominated by
questions of supply chain management, as is the case in the automotive industry.
This means that an MDO tool can be more impactful than in mature industries;
its outputs can be used directly for setting design specifications and
requirements.
Projected cost savings attained through design optimization can be realized
because their is no industry inertia to overcome; they don't require expensive
re-design of tooling, infrastructure (e.g. airports), or legislation (e.g.
Federal Aviation Regulations)\cite{leeham2018dont}.

\section{Geometric Programming:\\A Goldilocks Optimization Technique}
Due to the complex and inextricably coupled nature of the hyperloop design space, a
powerful, disciplined, and effective optimization tool is needed.
The rapidly iterative nature of engineering design, and our need to be
able to ``explore the design space,'' require that such a tool also be fast.

In optimization there is, broadly speaking, a tradeoff between model generality
-- how accurately an arbitrarily complicated design problem can be represented
-- and solver efficiency.
At one extreme, linear programs are a well-known class of convex optimization
problem that can be solved extremely quickly but are only valid for problems
that can be described with linear constraints.
At the other extreme, many general nonlinear local and global optimization
techniques have been developed for solving arbitrarily complex design problems,
but these techniques are often computationally intensive and provide no
guarantees about the quality of their solutions.

One goldilocks technique is called geometric programming\cite{boyd2007tutorial}.
Geometric programs (GPs) are able to model relatively complex relationships and
can be transformed into a convex optimization problem which can be solved
efficiently using off-the-shelf software that guarantees global optimality (if
a solution exists).
Crucially, these solvers also do not require an initial guess.
Being able to find optimal solutions without an initial guess makes the
technique particularly useful for the conceptual design of a transportation
system with no precedent.

These impressive properties are possible because GPs represent a
restricted subset of nonlinear optimization problems. In particular, the
objective and constraints can only be composed of \emph{monomial} and
\emph{posynomial} functions.

A monomial is a function of the form
\begin{equation}
m(\boldsymbol{x}) = c \prod_{j=1}^n x_j^{a_j},
\label{monomial}
\end{equation}
where $a_j \in \mathbb{R}$, $c \in \mathbb{R}_{++}$,
and $x_j \in \mathbb{R}_{++}$.
For instance, the familiar expression for lift, $\frac{1}{2}\rho V^2 C_LS$,
is a monomial with $\boldsymbol{x} = (\rho, V, C_L, S)$,
$c = 1/2$, and $\boldsymbol{a} = (1, 2, 1, 1)$.

A posynomial is a function of the form
\begin{equation}
\label{posynomial}
p(\boldsymbol{x}) = \sum_{k=1}^K c_k \prod_{j=1}^n x_j^{a_{jk}},
\end{equation}
where
$\boldsymbol{a_{k}} \in \mathbb{R}^n$,
$c_k \in \mathbb{R}_{++}$,
and $x_j \in \mathbb{R}_{++}$.
Thus, a posynomial is a sum of monomial terms, and all monomials are posynomials of just one term.

A GP minimizes a posynomial objective function, subject
to monomial equality constraints and posynomial inequality constraints. The
standard form of a geometric program is:
\begin{align}
\textnormal{minimize} \hspace{7pt} p_0(\boldsymbol{x}) & \nonumber \\
\label{gpdef}
\textnormal{subject to} \hspace{7pt} p_j(\boldsymbol{x}) & \leq 1, \hspace{5pt}
j = 1, ..., n_p, \\
 m_k(\boldsymbol{x}) & = 1, \hspace{5pt} k=1, ..., n_m, \nonumber
\end{align}
where the $p_j$ are posynomial (or monomial) functions, the $m_k$ are monomial
functions, and $\boldsymbol{x} \in \mathbb{R}^n_{++}$ are the decision
variables.

While this may seem restrictive, a perhaps-surprisingly broad range of physical
and economic relationships can be described using these sorts of relationships,
either exactly, through some algebraic manipulation or changes of variable; by
use of suitable approximations; or by fitting of GP-compatible surrogate
functions\cite{burnellphd,hoburg2016data}.

One simple and frequently used modeling technique that is central to the GP design
paradigm, known as the ``epigraph method''\cite{boyd2004convex}, is the
practice of relaxing a posynomial equality constraint into an inequality to make
it GP-compatible.
When the constraint is active, this relaxation is tight, meaning that equality
will hold at the optimum.

It turns out that it is possible to model a full hyperloop system to an
appropriate level of fidelity using geometric programming for every
relationship except for one which requires a \emph{signomial} constraint.

A signomial is a generalization of a posynomial that allows subtraction of
monomial terms, i.e. $c_k \in \mathbb{R}$.
Similarly, a signomial program is a generalization of a geometric program that
allows signomial constraints in addition to monomial and posynomial
constraints.
A signomial program can be solved by taking a local approximation of any
signomial constraints, solving the resulting geometric program, and repeating
this process until convergence is attained\cite{boyd2007tutorial}.

Signomial programs do not offer the same guarantee of global optimality as
geometric programs.
However, from a practical engineering perspective, when only one constraint is
signomial, and when
the range of possible values for such a variable is well understood, the
quality of solutions can be verified by plotting a
sweep over that variable.

Geometric and signomial programming have been used previously in a variety
of engineering applications, including the design of
aircraft\cite{hoburg2014geometric,kirschen2018application}, communication
systems\cite{chiang2005geometric}, and digital circuits\cite{boyd2005digital}.

A number of domain-specific languages have been written to empower engineers to
express design problems as convex optimization problems without needing to be
experts in the details of optimization algorithms
\cite{agrawal2019dgp,burnellphd}.
GPkit is one such modeling toolkit specifically designed for
formulating, manipulating, and solving geometric and signomial
programs\cite{burnell2020gpkit}.
It includes a number of powerful features such as automatic unit conversion,
model debugging, variable vectorization, solution ``diffs'', and a solution
post-processor that warns users if constraints are unexpectedly ``tight'' or
``loose''.
GPkit supports several backend solvers including MOSEK\cite{mosek} and
CVXOPT\cite{andersen2013cvxopt}.

\section{HOPS: The Hyperloop System Optimization Tool}
The remainder of this paper describes HOPS, the hyperloop system
optimization tool developed at Virgin Hyperloop\footnote{Virgin Hyperloop is
the latest name for a company that has previously also been known
as Hyperloop Technologies, Hyperloop One, and Virgin Hyperloop One.}.
From an engineering perspective, HOPS is a collection of coupled subsystem
models and a tool for quickly performing system-level trade studies and
sensitivity analyses.
It encompasses everything from the propulsion and levitation subsystems, to the
pod aerodynamics, to the pressure management pumps, to the size of the tube and
the number of podbays needed in each portal.

From a mathematical perspective, HOPS is a signomial program that is solved as
a sequence of geometric programs.
The default HOPS case at the time of writing has 4512 free variables and 6140
constraints, all of which are strictly GP-compatible (monomial or posynomial)
except for one important but simple signomial constraint.

From a software perspective, HOPS is a Python package that uses
GPkit as its modeling toolkit and MOSEK as its solver.
At its most distilled, HOPS encodes the optimization problem in three objects:
an objective function, a list of constraints, and a dictionary of input values
for any constants or variables whose values are fixed.
It takes a 2016 MacBook Pro with a sixth-generation Intel i5 processor
about 4 seconds to return an optimal solution for the default case.

The list of constraints is structured as a collection of models.
A description of the subsystem (and sub-subsystem) models follows, with
explanations of how the model was made GP-compatible, where appropriate.
A discussion of the relationship that needs a signomial constraint follows
as well.
But first, we address the most important question of an optimization problem:
what is the objective function?

\subsection{Objective function}
At a high-level, hyperloop system optimization can be thought of as a trade-off
between the cost to build a system, the cost to operate the system, and the
performance of the system, so it is appropriate to choose an objective function
that captures these three metrics.
The function we minimize is called the total cost per passenger-km.
\begin{align}
	\min \quad C_{\text{total, per pax-km}}
\end{align}
It is defined as the route-length-normalized sum of capital expenditure
(CapEx) per passenger, operating expenditure (OpEx) per passenger, and a
virtual time cost per passenger\footnote{We call the sum of CapEx per
passenger-km and OpEx per passenger-km the total \emph{hard} cost per
passenger-km or the Levelized Cost of Transportation (LCOT).  This is analogous
to a concept in power systems engineering called the Levelized Cost of Energy
(LCOE)\cite{ashuri2014multidisciplinary}.}.

\begin{align}
	C_{\text{total, per pax-km}} &= C_{\text{total, per pax}}/l_{\text{route}}\\
	C_{\text{total, per pax}} &= C_{\text{capex, per pax}} + C_{\text{opex, per pax}} + C_{\text{time, per pax}}
\end{align}

Whilst the route-length normalization is not strictly necessary for
single-route optimization, it has two advantages.
Firstly, it allows for easy comparison of solutions for different route lengths
as well as comparison against existing and competing modes of transportation,
for which the equivalent value can be calculated.
Secondly, it enables weighted optimization over multiple routes of different
lengths.

CapEx includes the cost to build the linear infrastructure, the portals and the
pressure management system, as well as the cost to manufacture the pod fleet
and is measured as an absolute cost (e.g. $\$$).
OpEx is the cost to operate and maintain the system, including both energy and
non-energy costs, and is measured as a cost-per-unit-time (e.g.
$\$/\text{year}$).
The time cost is a measure of the lost utility of a passenger's time -- the
opportunity cost of the time spent traveling -- and is measured as a
cost-per-passenger (e.g. $\$/\text{pax}$).
Estimating the cost of passenger time is a standard government practice for
evaluating the social benefit of transportation infrastructure
projects\cite{lee2000methods}.

To combine these three concepts into a single value, they must all
be normalized to units of cost-per-passenger ($\$/\text{pax}$).
To do this, both CapEx and Opex must be divided by the number of passengers
(i.e. trips) per year, $n_{\text{paxperyear}}$.
CapEx must also be multiplied by a parameter known as the capital recovery
factor, $f_{\text{cr}}$, which can be thought of as the inverse of a present
value (PV) factor, and is a function of the effective interest rate and project
payback period.
\begin{align}
	    C_{\text{capex, per pax}} &=  f_{\text{cr}} (C\!A\!P\!E\!X)_{\text{total}}/n_{\text{paxperyear}} \\
	    C_{\text{opex, per pax}} &=  (O\!P\!E\!X)_{\text{total}}/n_{\text{paxperyear}}
\end{align}
In its most simple form, the time cost per passenger is calculated by
multiplying travel time by a value of passenger time, $V_{\text{time}} [\$/hr]$, that
depends on how a population values productivity and
leisure\cite{small2012valuation}.
The Department of Transportation (DOT) periodically publishes an updated value
of passenger time for the United States\cite{white2016revised}.
The cost of passenger time calculation can be enhanced by also considering the
waiting time and using a wait-time multiplier, $M_{\text{waittime}}$, to quantify the
psychological phenomenon that makes time spent waiting less bearable than time
spent in motion\cite{wardman2004public}.
\begin{align}
	    C_{\text{time, per pax}} &=  V_{\text{time}} (t_{\text{travel}} + M_{\text{waittime}} t_{\text{wait, average}})
\end{align}

\begin{figure*}[ht]
        \centering
	\includegraphics[width=0.95\linewidth]{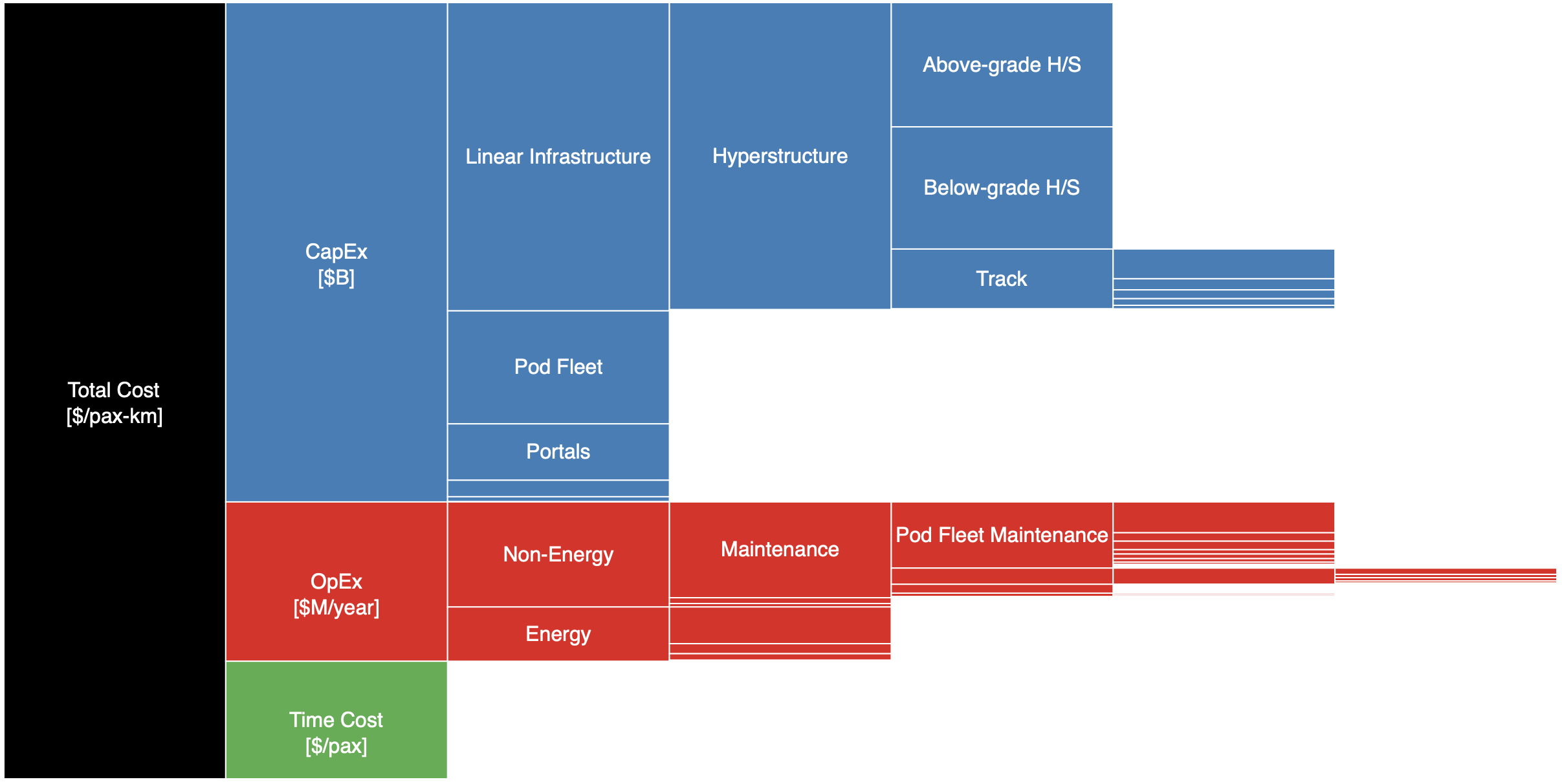}
	\caption{An icicle plot shows the breakdown of total cost for an
	example solution. By expressing all costs on a per-passenger-km basis,
	an apples-to-apples comparison can be made of capital cost, operating
	cost, and time cost.}
	\label{fig:icicle_example}
\end{figure*}

\autoref{fig:icicle_example} shows an icicle plot of the total cost breakdown
for an example solution.
The objective for HOPS is to shrink the total height of this chart as much as
possible, by minimizing and trading between the constituent cost buckets.

Although HOPS solves a single-objective optimization problem, it could be
thought of as a multi-objective optimization problem where the key weighting
parameters are the capital recovery factor, the cost of energy, and the value
of passenger time.

\subsection{Constraints}

Rather than being thought of as design-space limits, the constraints in HOPS
encode the relationships between all the variables.
They are organized into a hierarchy of models, where each model comprises three
elements: a declaration of variables (along with their respective units), a list
of constraints, and a dictionary of corresponding input values for variables
that are fixed\footnote{These inputs are referred to as ``substitutions'' in
GPkit verbiage to reflect the notion that any free variable can be substituted
with a fixed value.}, where applicable.
These models are constantly evolving, as constraints are added, removed,
simplified and replaced, and input values are updated, in the continuous
pursuit of representing the latest understanding of system and subsystem
characteristics and performance.

This pursuit is made more challenging by the tension between making models
accurate to the latest design point and capturing the sensitivities
and coupling of the broader design space.
The fidelity of models has generally increased over time but certain models
have also been simplified, for example when off-the-shelf hardware choices have
been made or when scaling laws have been deemed more appropriate than
first-principles physics models.
In addition, another key feature of HOPS is that different system and subsystem
architectures have been modeled, allowing discrete trade studies to be
performed.
As such, this section should be thought of as a glimpse at a snapshot of the
model at the time of writing, with the current default architecture.

\begin{figure*}[ht]
        \centering
	\includegraphics[width=\textwidth]{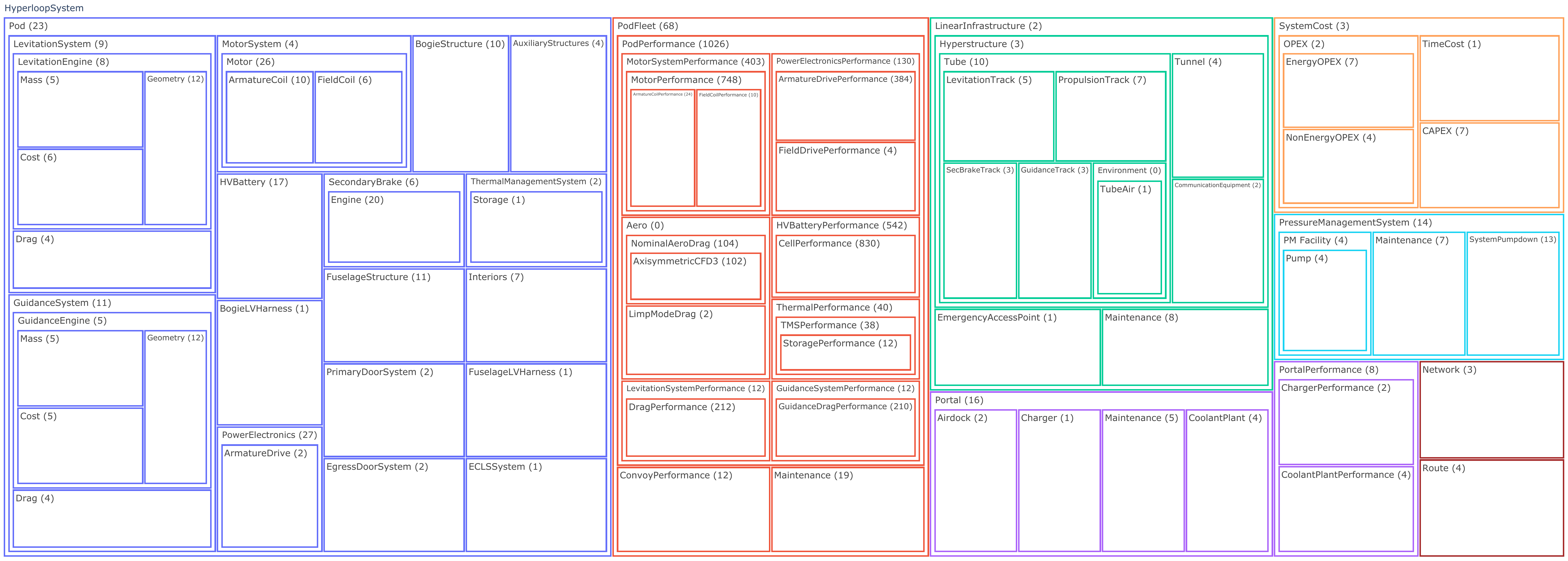}
	\caption{A treemap showing the HOPS model hierarchy for the default
	architecture, where every block represents a model and the number of
	constraints in each model is shown next to its name.  In total, there
	are 6140 constraints and 4512 free variables for the default case.}
	\label{fig:model_structure}
\end{figure*}

\autoref{fig:model_structure} shows the hierarchy of HOPS models.
There are models for each major system element (Pod, Linear Infrastructure,
Portal, and Pressure Management System) as well as models for the Route, Network,
and System Cost.
Most of these models have many levels of hierarchy of their own sub-models.

Although all models share a common structure and most have numerical
inputs, the Route and Network models can be thought of as the high-level
input models.
Similarly, the System Cost model can be thought of as the output model:
all other models feed into it and it encodes the cost rollups
that build up to the objective function.
That said, it is important to re-emphasize that HOPS is \emph{not} a
procedural design tool; it is fundamentally a list of constraints, \emph{not} a
sequence of calculations.
The added benefit of formulating the design as a convex optimization problem is
that it also offers a cleaner way to represent the recursive
relationships of a hyperloop system.

Note the separation of the Pod model
and the Pod Performance model (which sits inside the Pod Fleet model).
While it may seem redundant, this structure of modeling allows the
separation of variables (and constraints) that are route-invariant (e.g. pod
mass) and those that depend on the route(s) over which a pod is being
optimized (e.g. battery cell depth of discharge).
This in turn allows HOPS to be used in several different ways:
\begin{enumerate}
	\item Single-point optimization (nominal): Optimizing the system design
		and performance for a given route
	\item Fixed-design optimization: Optimizing the system for a sizing
		case, then freezing the design and optimizing only the
		performance for a different route
	\item Multi-point optimization: Optimizing a single system design over
		multiple routes each with an optimal performance
\end{enumerate}
Thanks to the design of GPkit, the separation of sizing constraints (Pod) and
performance constraints (Pod Performance) allows implementation of these
variants to be no more than a couple lines of code.
Perhaps counterintuitively, the Pod Fleet model is also a performance model
because the size of a fleet can vary significantly for different routes and
business cases, and is necessarily a key optimization variable.

The remainder of this section provides an overview of each top-level model,
along with more detailed descriptions of a few key sub-models, where
appropriate.
The purpose is to demonstrate the breadth and depth of models, and in doing so,
give the reader a sense for HOPS' level of fidelity, which ranges from fairly
crude for the Route model to quite detailed for key subsystems of the Pod
model.

\subsubsection{Route}
The Route model encodes the high-level inputs for an origin-destination pair.
This includes a crude representation of the alignment (i.e. the path of
the linear infrastructure) and the expected passenger traffic.
Key inputs include the route length, the net elevation
change, the maximum grade, the fraction of the alignment that is below grade
(tunneled), a parameter that describes the straightness of a route (discussed
further in \autoref{sec:convoy}), and the peak and average hourly passenger
throughput.
In the current model, throughput is taken as fixed and does not
depend on travel time;
however it can easily be adapted to include a demand response to performance.

\subsubsection{Network}
A key part of the hyperloop value proposition is the ability to offer
direct-to-destination service, a feature that requires on- and off-ramps, much
like a highway.
\begin{figure}[ht]
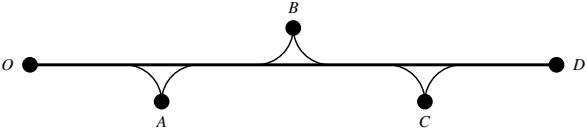

\begin{center}
	\includestandalone[scale=0.7]{figs/route_topology}
	\caption{Schematic of network topology for an example network with five
	portals. A mainline connects the terminus portals (O and D), while
	ramps serve the bypass portals (A, B, and C).}
\end{center}
\end{figure}
While certain networks will allow local reduced speed in the vicinity of
portals, the only general solution that does not adversely impact
throughput capability is for the mainline to have a constant
cruise velocity.
Therefore, ramps need to provide enough distance to accelerate
to full cruise velocity.
The length of these ramps is determined by the acceleration and deceleration
capabilities of pods, and in turn directly impact the total length, and
thus cost, of linear infrastructure that needs to be built.
The total network length is captured as a simple-but-critical posynomial
constraint:
\begin{align}
	l_{\text{network}} \geq l_{\text{longestroute}} + n_{\text{ramps}} l_{\text{ramp}}
\end{align}
For the purpose of this work, only single-origin-destination-pair networks are
considered.
Because it does not need ramps, optimizing around such a network would yield
a system with low acceleration capability and thus limited long-term
value proposition, a parameter can be used to represent the number of
``extra'' ramps, as a surrogate for a more sophisticated implementation with
multiple explicitly-defined origin-destination pairs.
Discussion of this more sophisticated implementation is left for future work.

\subsubsection{Pod and Pod Fleet}
The pod is the most complex and multidisciplinary system element and the most
ripe for optimization.
As such, the models that describe its design and performance are the most
detailed, and its system-level specifications are amongst the most important
products of HOPS.
The high-level sizing variables include the pod mass, the pod capacity (i.e.
number of passengers), the pod cost, and the variables that describe its
geometry.
The key performance variables for a given route are those that describe the
velocity profile (e.g. launch acceleration and cruise velocity), travel time,
energy consumption, and thermal losses.

The Pod model comprises sizing models for each pod subsystem.
Each sizing model captures the mass and cost of that subsystem, using a
combination of roll-up constraints (e.g. $m \geq m_1 + m_2 + ...$), scaling law
constraints, and input values.
Some subsystems also have models that capture their geometry
(e.g. volume, frontal area, length, height, and width) and these variables tie
into constraints that capture considerations such as aerodynamic drag, pod structure
sizing, tube sizing, and portal sizing.
Key active subsystems have corresponding performance models (under Pod Performance)
that capture maximum power and total energy consumption.
Subsystems with significant losses also include performance models for (instantaneous)
thermal power, (aggregated) thermal energy, and drag, where applicable.
There is also a simple model to account for the maintenance costs of each
subsystem.

As \autoref{fig:model_structure} shows, there are 16 major pod subsystem models,
however only 7 of these are modeled to a level of fidelity where they have
significant direct coupling to other subsystems.
These are the High-Voltage Battery, Power Electronics, Propulsion, Levitation,
Guidance, Secondary Brake, and Thermal Management Systems.
All of these, with the exception of the Secondary Brake, also have
corresponding performance models.
Two examples of these subsystem models are highlighted below to show the level
of fidelity considered and to demonstrate how modeling tricks are able to
make a relatively complex power loss model GP-compatible.

There is also a performance model to capture the aerodynamic drag
experienced by the pod throughout its trajectory.
We highlight it below to show how GP-compatible fits can be used to
capture complex non-analytical design spaces.
We also describe the model vectorization that allows us to account for the
different aerodynamic drag experienced by different pods in a convoy.

Pods do not operate in isolation; they are part of a fleet and they are capable
of traveling in convoys.
The constraints that model the fleet are in the top-level Pod Fleet model.
The number of pods in the fleet for a given route is a key free variable
because the cost of the fleet represents a significant proportion of total
CapEx and therefore total cost per passenger-km.
The size of the fleet depends on the peak throughput for the route, the
pod capacity, the travel time and the in-portal turnaround time.
Because the turnaround time is typically constrained by the battery charging
time (with current battery technology) for all but the shortest routes, the
fleet size is therefore coupled not only to the travel time but also to the
energy efficiency of the pods.

Not all pods in a convoy have the same
energy consumption, but they must all follow the same velocity
profile.
As such, the critically important constraints that model travel time and the
velocity profile, including the only non-GP constraint in HOPS, are part of
the Convoy Performance model.

\begin{figure}[ht]
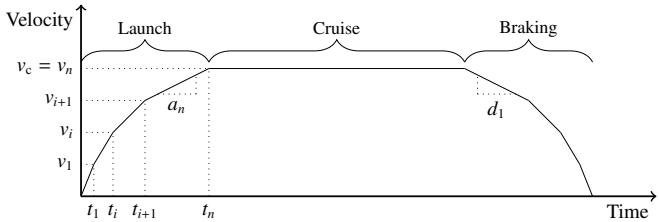

        \centering
	\includestandalone[scale=0.85]{figs/velocity_profile}
	\caption{The nominal velocity profile modeled in HOPS has a launch
	phase, a cruise phase and a braking phase. The launch and braking
	phases are discretized into $n$ equal-velocity steps.}
\label{fig:velocity_profile}
\end{figure}

\paragraph{Convoy Performance}
\label{sec:convoy}
The velocity profile model underpins all of the other pod performance models and
is therefore critical to system optimization.
The trajectory modeled in HOPS has a launch (acceleration) phase, a
constant-speed cruise phase, and a braking (deceleration) phase.
The launch and braking phases are discretized into a user-specified number of
steps (default value: 25) to improve the accuracy of the model, for example by
allowing the acceleration rate to vary which is particularly important for
capturing the behavior of the system when it is power-limited.
This means that all launch and braking constraints are ``vectorized'' (i.e.
duplicated for every step) so every launch or braking constraint actually
represents 25 constraints.
Thanks to a GPkit feature called Vector Variables, each constraint can just be
written once and can mix Vector Variables with (scalar) Variables.
For example, the constraint derived from Newton's second law
\begin{align}
	\mathbf{F_{launch}} \geq m_{\text{pod}} \mathbf{a_{launch}} + \mathbf{D_{launch}}
\end{align}
actually represents
\begin{align}
	F_{\text{launch},0} &\geq m_{\text{pod}} a_{\text{launch},0} + D_{\text{launch},0}\\
	F_{\text{launch},1} &\geq m_{\text{pod}} a_{\text{launch},1} + D_{\text{launch},1}\\
	&...\nonumber\\
	F_{\text{launch},24} &\geq m_{\text{pod}} a_{\text{launch},24} + D_{\text{launch},24}
\end{align}
To ensure a GP-compatible formulation, this discretization is performed using
uniform velocity steps (instead of uniform time steps) as illustrated in
\autoref{fig:velocity_profile}.

The travel time model is important not only because of the time cost component
of the objective function, but also because travel time directly impacts the
size of the fleet needed for a given route.

The velocity-based discretization allows us to model travel time using the
following posynomial constraint, where $l_{\text{route}}$ is the total route length,
$v_\text{c}$ is the cruise velocity, $n_i$ is the number of discretizations of the
launch phase, $a_i$ is the acceleration rate during each launch segment, $n_j$
is the number of discretizations of the braking phase, and $d_j$ is the
deceleration rate during each braking segment.
\begin{align}
	t_{\text{travel}} &\geq \frac{l_{\text{route}}}{v_\text{c}} + \frac{v_\text{c}}{2}\left(
        \sum_{i=1}^{n_i} \frac{1}{a_i} \left(\frac{2n_i - 2i + 1}{n_i^2}\right) +
        \sum_{j=1}^{n_j} \frac{1}{d_j} \left(\frac{2j - 1}{n_j^2}\right)\right)
\end{align}
Although at first glance, this may not appear GP-compatible due to the
negative terms, these terms do not include any model variables and simplify to
give a posynomial expression.

As mentioned previously, there is one signomial constraint in HOPS.
This is the constraint that governs cruise length, which must have a lower
bound constraint because other models apply downward pressure on both cruise
length and cruise time:
\begin{align}
	f_{\text{cruise}} \geq 1 - 2 l_{\text{ramp}}/l_{\text{route}}
\end{align}
where $f_{\text{cruise}}$ is the fraction of the trajectory in cruise and
$l_{\text{ramp}}$ is the ramp length.
We know this constraint is not GP-compatible by inspection because its right-
hand side is an example of the ``one-minus'' atom, which is log-log
concave\cite{agrawal2019dgp}.
There are many possible formulations of this constraint that would achieve the
same intent but they all generally have the same structure.
It is also possible that a clever GP-compatible formulation exists but has not
yet been derived.
Note also that if the value of ramp length is fixed or a sweep of ramp length is
conducted, then this constraint is no longer a signomial and HOPS can be solved
with a single GP, rather than a sequence of GPs.

Unfortunately, not all of the alignments proposed for prospective routes are
able to support a constant-cruise-velocity profile at the velocities a
hyperloop system is capable of, due to curvature and passenger comfort
constraints.
HOPS is not able to optimize over arbitrarily complex velocity profiles
but optimizing over an ideal trajectory can significantly underestimate the
electrical, thermal, and travel time impacts of more complex routes and
limiting the cruise velocity throughout the trajectory would be unrealistically
conservative.
It is therefore necessary to have a way to account for these effects.

\begin{figure}[ht]
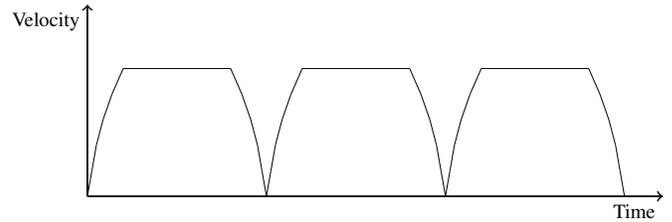

        \centering
	\includestandalone[scale=0.85]{figs/velocity_profile_slowdowns}
	\caption{Velocity profile with two full slowdowns}
\label{fig:velocity_profile_slowdowns}
\end{figure}

We use a parameter known as the ``number of full slowdowns'' for this purpose,
where a full slowdown is defined as a deceleration to rest followed by a
re-acceleration to cruise velocity.
An appropriate value for this parameter is provided by an in-house simulation
tool that can determine the equivalent number of full slowdowns for an arbitrary
velocity profile.
\autoref{fig:velocity_profile_slowdowns} shows an example velocity profile with
two full slowdowns, though it should be noted that non-integer values can be
used.

\paragraph{Battery}

The size of the battery\footnote{Battery here refers to the high-voltage
battery for the sake of succinctness. There is also a low-voltage battery that
is much less significant to the design.} is one of the most important design
decisions for a podside-powered hyperloop system.
Not only does the battery significantly impact key performance metrics like
acceleration capability and nonstop range, it is also the heaviest
subsystem, the largest subsystem inside the bogie structure, and the
biggest contributor of thermal energy, so it has a significant impact on both
the design of other major subsystems (e.g. levitation, propulsion, thermal
management) and the overall size of the pod structure (and in turn, the tube
structure).

As with other major subsystems, the battery is also subject to critical recursive
design relationships.
If it gets heavier, the whole pod gets heavier, which then
requires the battery and other key subsystems to be more powerful (i.e.
heavier) to achieve the same performance.
Such mass spirals can be caused by any change to the major subsystems.

\subparagraph{Design Considerations}

\begin{figure}[ht]
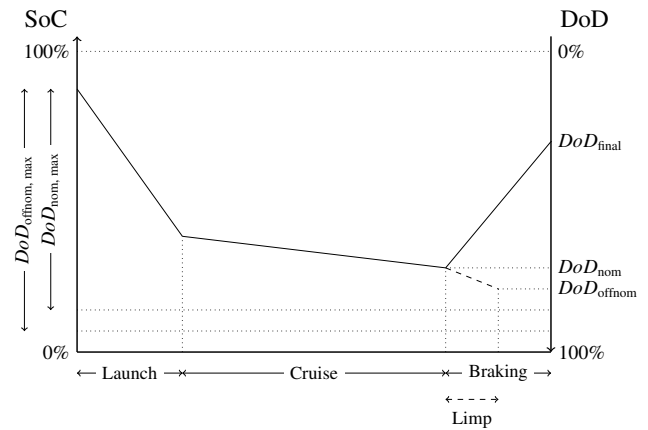

	\centering
	\includestandalone[scale=0.7]{figs/battery_soc}
	\caption{Illustration of variables that pertain to depth of discharge
	(DoD) and state of charge (SOC). The nominal DoD and off-nominal DoD
	are bounded by the maximum nominal DoD and maximum off-nominal DoD,
	respectively. The final DoD determines how much energy must be
	recharged in portals.}
	\label{fig:battery_soc}
\end{figure}

The battery must be able to supply the power demanded of it by other subsystems
in all phases of a trajectory and it must store enough energy to complete the
journey in both nominal and off-nominal scenarios.
Because the system uses regenerative braking, this energy constraint applies to
the lowest depth-of-discharge point of a route, which occurs at the end of
cruise for a nominal trajectory, as shown in \autoref{fig:battery_soc}.

These requirements must be met whilst also respecting important design
constraints.
First, the battery must operate within the current
and voltage capabilities of the cells, during all phases.
These limits can come from cell manufacturer datasheets as well as test results
or other design considerations.
Second, heat generated by the battery must be absorbed, or otherwise
dissipated, by the thermal management system (TMS).
This can be thought of as a bilateral constraint on both the battery and the
TMS (i.e. should the battery generate less heat or should the TMS get bigger?).
Third, cells need to be replaced when their performance deteriorates to a
certain point, but cell replacement costs can be a significant portion of system
OpEx.
The battery replacement rate is therefore another key operating
decision, trading off replacement cost with the capacity, efficiency and
thermal load impacts of using cells for longer.
Finally, battery charging time is often (though not always) the gating
factor for pod turnaround time, which directly impacts the size (and therefore
cost) of portals.

\subparagraph{Modeling Fidelity}

The battery is modeled using basic cell physics, including the heat
generated by, and voltage drop due to, ohmic losses.
These constraints are duplicated for every phase of a trajectory,
including each discretization step of the launch and braking phases:
\begin{align}
	P_\text{cell,cruise} + P_\text{loss,cell,cruise} &\leq
		V_\text{opencircuit,cell} I_\text{cell,cruise} \\
	P_\text{cell,cruise} &= V_\text{terminal,cell,cruise} I_\text{cell,cruise} \\
	P_\text{loss,cell,cruise} &\geq I_\text{cell,cruise}^{2} (R_\text{internal,cell,cruise}
		 + R_\text{contact,cell})
\end{align}

\begin{figure}[ht]
	\centering
	\includegraphics[width=0.9\linewidth]{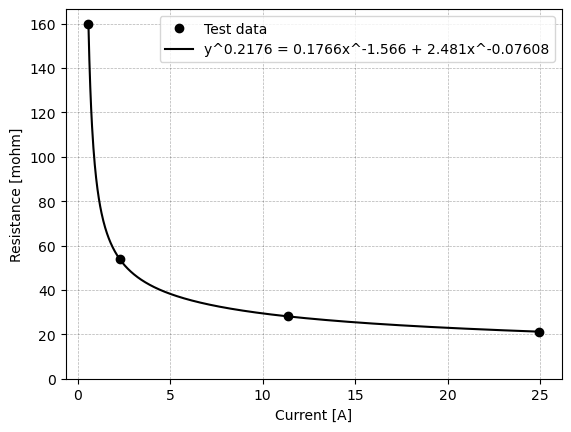}
	\caption{GP-compatible fit of test data for cell internal resistance as
	a function of discharge current}
	\label{fig:murata_fit}
\end{figure}

The model also includes constraints that capture the effect of cell aging on
internal resistance and capacity, using GP-compatible fits of empirical data.
An example of a fit of the internal resistance of a cell as a function of
discharge current is shown in \autoref{fig:murata_fit}.

Cell properties such as beginning-of-life capacity and maximum charge and
discharge current limits are inputs to the model and different sets of inputs
can be selected to represent different cell options and perform trade studies.
Input sets can also be used to study the impact of potential future
improvements in cell technology.

\paragraph{Propulsion System}
\label{sec:propulsionsystem}

The proprietary propulsion system\cite{jedinger2021homopolar} consists of
on-board linear motors that interact electromagnetically with a passive wayside
track and provide bi-directional longitudinal forces, enabling the pod to
accelerate, overcome drag, and climb grades, while also providing nominal
regenerative braking.

The capability of the motors directly impacts the pod's acceleration and
deceleration performance, and therefore the length of ramps.
As with most motors, under full load they operate in two regimes: constant
force and constant power.
The speed at which the transition from constant force to constant power occurs
is called the base speed.

\subparagraph{Design Considerations}
The propulsion system must be capable of delivering (or absorbing) the force
and power required in every phase of a trajectory.
In the launch and cruise phases, this force includes the drag, any
acceleration, and any grade climbing.
The total drag is the sum of the aerodynamic drag, the drag due to the
levitation and guidance systems, and, perhaps confusingly, the drag due to the
motors themselves because of losses in the track.

The losses generated by the motors are another major design consideration, both
from the perspective of sizing the power electronics and the battery, and
sizing the thermal management system.
To improve the overall efficiency of the propulsion drive train (including the
power electronics and battery) in cruise, a parameter allows use of a subset of
the motors during low-power operation, so that the power electronics can be
operated more efficiently.

As with every other subsystem, we must also consider the mass, cost, and
geometry of the motors.

Another significant design consideration is the cost of the propulsion track,
which depends on the cross-sectional area of the track and therefore on the
size of the motors.

\subparagraph{Modeling Fidelity}
The motor is modeled using a combination of basic physics models and geometric
scaling laws.
For example, motor force scales with the active area, using an input for
force-per-unit-area taken from detailed electromagnetic simulation work, and
validated through testing.
The basic geometry of every key component (e.g. coils, cold plates) in the
motor is modeled and the motor active width is the most significant sizing
variable.
Some design decisions like coil pitch are made using more detailed
electromagnetic analysis and taken as inputs to HOPS.

The various sources of loss both on-board and in the track are modeled in
relative detail.
The losses in the coils are modeled as functions of current and resistance,
where the resistances of the coils are modeled as functions of geometry.
Other on-board losses are modeled as a function of flux and motor geometry.
Track-side losses are modeled as a function of motor geometry, pod velocity,
and flux.
Accurately capturing the loss models is made more complicated because
different relationships apply to the constant-force and constant-power phases of
acceleration.
To keep relationships GP-compatible, the maximum power loss and the energy loss
(i.e. the integral of power loss) must therefore be modeled separately.
Several GP-compatible fits are necessary to make this model work, including
some fits of the integral of certain variables.

\begin{figure}[ht]
        \centering
	\includegraphics[width=0.9\linewidth]{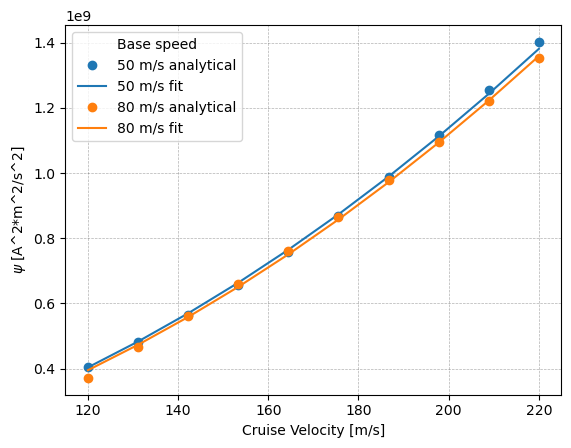}
	\caption{A GP-compatible fit of the term in square brackets as a
	function of cruise velocity and base speed over a likely range of
	values. The function is weakly dependent on the base speed over the
	domain of interest.}
	\label{fig:integralfit}
\end{figure}

For example, the total loss in one of the coils during launch is calculated as
the integral of power over the time reach cruise:
\begin{align}
	E_\text{loss,coil,launch} = \int_0^{t_\text{c}} P_\text{loss,coil} dt = 3 R_\text{coil} \int_0^{t_\text{c}} I_\text{coil}^2 dt
\end{align}
But $I_\text{coil}$ is a piecewise function of velocity with different behavior
either side of the base speed $v_\text{b}$, i.e. in the constant-force and
constant-power phases:
\begin{align}
        I_\text{coil}^2 &=
\begin{cases}
        I_\text{rated}^2 & v \leq v_\text{b} \\
        I_\text{rated}^2 \left(\frac{v_\text{b}}{v}\right)^2 + I_\text{d0}^2 \left(1 - \frac{v_\text{b}}{v}\right)^2& v > v_\text{b} \\
\end{cases}
\end{align}
so it needs to be integrated piecewise.
\begin{align}
	\int_0^{t_c} I_\text{coil}^2 dt &= \int_0^{t_b} I_\text{rated}^2 dt
	  + \int_{t_b}^{t_c} I_\text{rated}^2 \left(\frac{v_\text{b}}{v}\right)^2
	  + I_\text{d0}^2 \left(1 - \frac{v_\text{b}}{v}\right)^2 dt\\
		   &= I_\text{rated}^2 \frac{v_\text{b}}{a_0} + \frac{m_\text{pod}}{P_\text{peak}} \Biggl[
			   \left(I_\text{rated}^2+I_\text{d0}^2\right) v_\text{b}^2 \log\left(\frac{v_\text{c}}{v_\text{b}}\right) \\
		   &+ I_\text{d0}^2\left(\frac12 v_\text{c}^2 - 2 v_\text{b} v_\text{c} + \frac32 v_\text{b}^2 \right) \Biggr] \nonumber
\end{align}
The cumbersome expression in square brackets is not GP-compatible, but we can
approximate it well using a posynomial fit with 2\% RMS error as shown in
\autoref{fig:integralfit}.

We can now express the loss with the following constraint, where $\psi$ is the
posynomial function plotted in \autoref{fig:integralfit}.
\begin{align}
        E_\text{loss,coil,launch} &\geq 3 R_\text{coil} \left(I_\text{rated}^2 \frac{v_\text{b}}{a_0} + \frac{m_\text{pod}}{P_\text{peak}} \psi \right)\\
	\psi^{0.151} &\geq 2.34 v_\text{c}^{0.305} v_\text{b}^{-0.00877} + 2.34 v_\text{c}^{0.308} v_\text{b}^{-0.00328}
\end{align}

\paragraph{Aerodynamic Drag}
\label{sec:aero}
Although a defining feature of a hyperloop is a low pressure environment, there
is still enough air in the tube to cause aerodynamic drag that, while small,
cannot be neglected.

\begin{figure}[ht]
        \centering
        \includegraphics[width=0.9\linewidth]{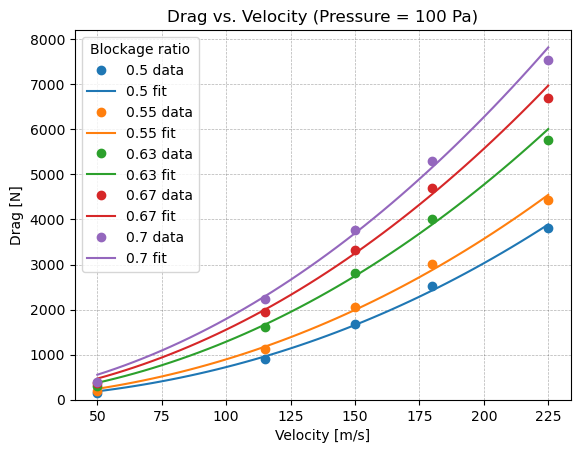}
	\caption{Slices of 3D surrogate model at different blockage ratios with
	a pressure of 100 Pa and a baseline pod frontal area of 11.4 $\text{m}^2$.}
	\label{fig:aeroslices}
\end{figure}

The drag coefficient depends primarily on three factors: the velocity of a pod,
the tube pressure, and the blockage ratio (the ratio of pod frontal area to
tube cross-sectional area).
In certain high-traffic scenarios, the aerodynamic drag can also depend on the
convoy-to-convoy spacing.
We use a GP-compatible fit of CFD data to build surrogate models that take some
or all of these dimensions into account.
\autoref{fig:aeroslices} shows slices of one such model (RMS error: 4\%).

A key feature of hyperloop operations is the ability for pods to travel in
convoys, much like a train but without any physical coupling between pods.
This enables a passenger throughput capacity that eclipses those of the best
high-speed rail systems in the world, while still supporting
direct-to-destination service.
However, convoy operations also offer another benefit: reduced aerodynamic
drag when averaged across all pods thus reducing overall system energy
consumption.

The average part is important; different pods in a convoy experience different
levels of aerodynamic drag.
CFD analysis has shown that, perhaps counterintuitively, the last pod in a
convoy feels the highest level of drag, approximately double the average drag
for a 4-pod convoy.
This difference is significant enough to justify modeling the two cases
separately and simultaneously.
While special ``caboose'' pods could be designed (with slightly larger
batteries and thermal management systems) this would be challenging from an
operational flexibility perspective, particularly when pods need to be able to
peel off from convoys to reach different destinations.
A single pod design therefore needs to be able to complete a given journey
regardless of its position in a convoy.
The pod should be optimized for the weighted-average performance but capable of
performing the most challenging case.
This means the last pod in a convoy is the ``sizing'' case.

GPkit allows this functionality to be implemented with a single command to
``Vectorize'' the Pod Performance model.
This duplicates all of the constraints and each Pod Performance variable now
has a two-element solution, one for the lead pods in a convoy, which all
experience similar levels of drag, and one for the last pod\footnote{The
dimensionality also scales with the number of routes. For example, when
optimizing over three routes, each Pod Performance variable has six elements.}.
Because the Pod Performance models represent a significant proportion of the
constraints in HOPS, this duplication significantly increases the size of the
problem, however thanks to the efficiency of the solvers used, the increased
computational cost is low in practical terms; the difference between 2 and 4
seconds for a typical point-solution.

The difference in the solution is significant enough to be worthwhile.
For typical solutions with a 4-pod convoy, there is a 10-20\% difference
between the average energy consumed and the energy consumed by the last pod,
which translates to a non-negligible difference in the system design, compared
to what it would be if only optimized for the last pod performance.

Importantly, the model can also still be run in ``lone-pod'' mode, with a
single Pod Performance model, for scenarios where convoys are less relevant
(e.g. low traffic routes).

This model vectorization could also be extended to other domains of the model
where it is helpful to optimize over a distribution of cases within a fleet,
such as different battery ages and different payload mass.

\subsubsection{Linear Infrastructure}
The linear infrastructure, or hyperstructure\footnote{The hyperstructure name
derives not only from hyperloop, but also from the generalization of
the substructure (i.e. columns) and superstructure (i.e. tube).}, provides an
enclosed environment through which pods travel, while supporting the track
elements needed for propulsion, levitation, guidance, and emergency braking.
Most routes will likely have an alignment with both above-grade (i.e.
elevated) and below-grade (i.e. tunneled) portions, which requires two
different hyperstructure designs.
The above-grade hyperstructure comprises a substructure (i.e. columns)
and a superstructure (i.e. tube); the below-grade hyperstructure comprises a
tube that can be constructed as part of the tunnel boring process.
HOPS does not perform alignment optimization, so the fraction of a route that
is below-grade is taken as an input.

The predominant consideration for the hyperstructure is cost and the
hyperstructure CapEx is often the largest single contributor to the total cost
per passenger-km.
As discussed previously, for a network with multiple ramps, the total length of
infrastructure is a significant driver of total infrastructure cost.
However, the cost of each unit-length of hyperstructure is also important and
depends on the cost of the civil infrastructure (e.g. columns, tunnel boring
etc.), the hyperloop-specific infrastructure (tube, tunnel liner etc.), the
track elements, and any wayside communication equipment.

The cost of the tube and the track elements are directly coupled to the design
of the pod.
There are two main considerations for tube size: geometric constraints and
aerodynamic blockage.
The tube must be large enough to fit the pod with appropriate
range-of-motion envelopes and to prevent significant aerodynamic drag, as
discussed in \autoref{sec:aero}.
The blockage ratio, $r_\text{blockage}$, also must consider the cross-sectional area
of track elements inside the tube.

\begin{align}
            r_\text{blockage} &\geq  A_\text{pod,frontal}/A_\text{tube,unobstructed} \\
            A_\text{tube,unobstructed} + A_\text{track,frontal} &\leq  A_\text{tube}
\end{align}

The cost of each track element depends on its size, which is directly coupled
to the geometry of the engine with which it interacts, and therefore the pod
mass and performance.
Track cost also depends on whether the track is laminated or solid, which also
affects the performance of the engines.

An important caveat for modeling the hyperstructure is that a large
portion of the cost is very insensitive to, or even independent of, the design.
This is in part due to the relatively (and necessarily) crude nature of
infrastructure design and construction, and in part because a lot of
infrastructure cost has nothing to do with engineering at all\cite{levy2020low}.
These large fixed costs, which can also be heavily project-dependent, must be
accounted for, not only in pursuit of accuracy for total cost, but also to be
able to communicate total cost sensitivities.
Understanding how the optimal system design looks under a variety of
infrastructure cost scenarios is important to producing a valid and defensible
design.

\subsubsection{Pressure Management System}
The Pressure Management System (PMS) has two main functions: maintaining the
operating tube pressure and pumping the system down to operating pressure after
maintenance or any other required re-pressurization of the tube.

Maintaining operating pressure involves pumping out any air that leaks into the
system during the course of normal operations.
Air leaks into the tube both from outside the tube and from inside the pods, so
the capacity of the PMS is coupled not only to the length of linear
infrastructure, but also to the number of pods in the system.

For pumpdown, the PMS must be able to return the system to nominal operating
pressure in a suitably short period of time to allow overnight maintenance
operations.
Pressure management facilities are collocated with gate valves so that sections
of the tube can be isolated for re-pressurization.
Therefore pumpdown requirements constrain not only the number of pumps but also
the spacing of pump facilities.

The key design variable for the PMS is the tube pressure.
Other important design variables include the total number of pumps and the
spacing of the pump facilities.
In addition to the capital cost associated with the pumps and the supporting
equipment needed at every pressure management facility, the PMS operating costs
must be accounted for.
These include the energy consumption for both pressure maintenance and pumpdown
operations, as well as pump maintenance costs.

\begin{figure}[ht]
	\centering
	\includegraphics[width=0.9\linewidth]{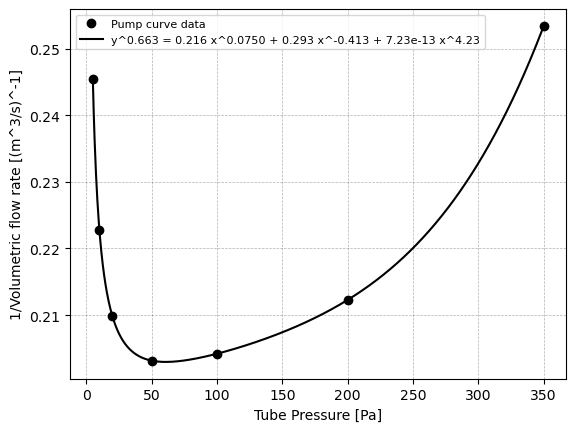}
	\caption{GP-compatible fit of pump curve data: flow rate as a
	function of pressure. Because there is upward (optimization) pressure
	on flow rate, we fit to the reciprocal of flow rate.}
	\label{fig:pumpcurve}
\end{figure}

The performance of the pumps is captured using GP-compatible fits of pump
manufacturer data for flow rate and power consumption as a function of pressure.
A fit for flow rate is shown in \autoref{fig:pumpcurve}.
Different pumps are modeled to allow comparisons of different technologies.

\subsubsection{Portals}
A portal can be decomposed into technology-agnostic infrastructure (i.e. a
building or underground structure), hyperloop-specific infrastructure (e.g.
tubes), hyperloop-specific hardware (e.g. battery chargers, coolant supply
equipment, and airdocks), and other equipment.
Each portal has a number of podbays where passengers board and
disembark while pods are charged and replenished with coolant.
Larger portals may have multiple branches of podbays to help with traffic
flow.

The vast majority of system energy consumption occurs in the portal, because
that is where pod batteries are charged and coolant is produced.
Passengers board pods through airdocks, gateways that provide a pressurized
connection between the pod interior and the platform.
Each time a pod departs, the air must be removed from the airdock, which also
has an associated energy consumption.

The size of each portal depends on the size and number of podbays needed to
serve the pod traffic during peak hour.
The size of podbays depends on the size of pods, and the number of
podbays depends on how long it takes for a pod to turn around, which is
constrained by both the time it takes to charge the battery and the time it
takes for passenger loading and unloading.
The time needed to charge the battery depends on how much energy the pod
consumed during its inbound journey.

As with the hyperstructure, portals have a large heavily-site-dependent fixed
cost component and significant cost uncertainty due to the challenges
associated with large infrastructure projects.

\section{Use of HOPS}

HOPS has four main uses: system sizing, performance optimization, sensitivity
studies (``sweeps''), and trade studies.

\subsection{System Sizing}
The primary purpose of HOPS is to propose optimal designs, given the best
available understanding of design relationships and input values.

A full HOPS solution describes everything from the total pod mass and
the tube diameter down to the number of cells in the battery, the mass of
coolant in the thermal management system, and the key dimensions of each
engine.

One of the key products of HOPS is the pod mass budget, which provides all pod
subsystem designers with a mass allocation.
Following standard system engineering practice, we use mass margins in HOPS to
account for unforeseen but likely increases beyond the current best
estimate.
This helps subsystem designers design around what the anticipated final pod
mass will be, rather than what it is at the time of solution \cite{robustexp,meluso2020}.

\begin{figure}[ht]
        \centering
        \includegraphics[width=\linewidth]{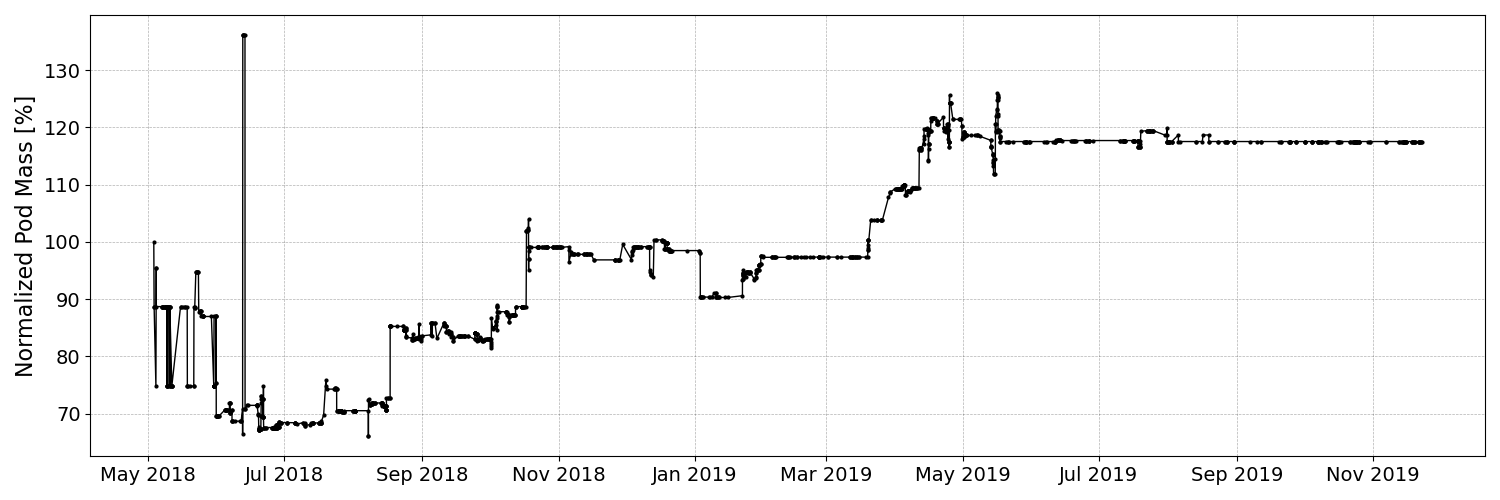}
	\caption{Pod mass against time for the first design cycle. Each point
	represents a commit to the HOPS code repository and the data is
	normalized by the first point.}
	\label{fig:history}
\end{figure}

Because the aforementioned ``best available'' understanding is constantly evolving
in the fast-paced environment of a startup, the
optimal solution is also always evolving.
This allows the implications of changes in one subsystem to be rapidly
propagated to all other subsystems.
\autoref{fig:history} shows how the optimal pod mass for the sizing case
evolved for the first design cycle, both as the design understanding improved
and as business requirements changed.

\subsection{Performance Optimization}
In addition to the sizing variables, every HOPS solution also describes the
optimal performance of the system over the route(s) in question.
This includes the optimal velocity profile, tube pressure, and pod turnaround
time, as well as low-level quantities like the current in the battery cells and
motor coils throughout the pod's trajectory.

The Pod and Pod Performance models are separated to allow for a one-line
freezing of all pod sizing variables.
This allows the pod design to be optimized for a particular sizing case and
then frozen for a subsequent solution over a different route to optimize its
performance for that route with the given design.

This capability is useful when the most business-relevant use-case differs from
the sizing case.
For instance, it is valuable to understand how a trade study impacts the total
cost per passenger-km for a near-term project opportunity, in addition to
understanding the impact on a challenging but more distant sizing case.

This capability is also particularly useful for business development.
Optimal costs, travel time, and energy consumption for a particular route
can be shared with prospective customers for the
purpose of feasibility studies and technology comparisons.

\subsection{Sensitivity Studies}
Interior point solvers return dual variables at no extra cost, so every HOPS
solution also includes the local sensitivity of the objective function to each
input value and constraint at the optimal point.
In addition to these, it can be useful and illustrative to explore the broader
design space with sweeps of inputs to see, for example, when certain
constraints become active.
Because HOPS solves a single design point in just a few seconds, performing
even large and multi-dimensional sweeps is relatively quick.
This allows system designers to understand the sensitivity to any input, from
low-level variables like motor loss model parameters or the leak rate of the
tube, to high-level variables like the route length or value of passenger time.

\begin{figure}[ht]
        \centering
        \includegraphics[width=0.9\linewidth]{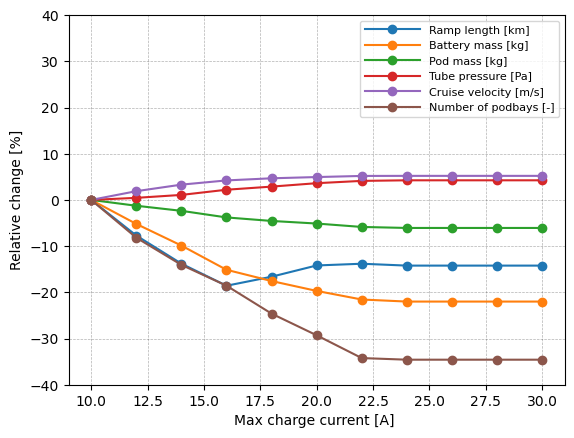}
	\caption{The effect of sweeping maximum charge current on six key
	variables, shown as percentage changes from the first point.}
	\label{fig:examplesweep}
\end{figure}

\begin{figure}[ht]
        \centering
        \includegraphics[width=0.9\linewidth]{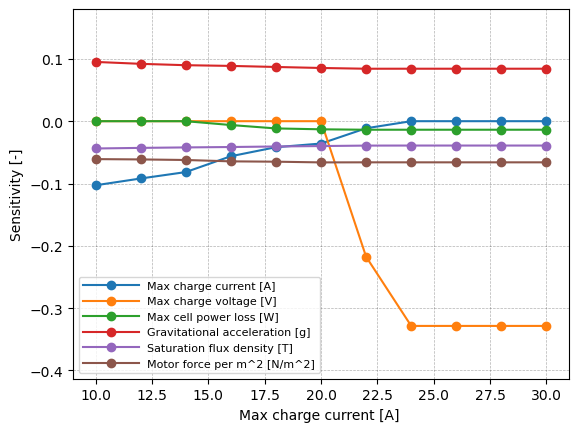}
	\caption{Changes in sensitivity help us to identify when constraints
	become active in the sweep shown in \autoref{fig:examplesweep}. The
	vast majority of inputs have a flat or shallow slope. A sensitivity of
	+1 means that decreasing that value by 1\% will decrease (improve) the
	objective function by 1\%.}
	\label{fig:sensplot}
\end{figure}

For example, \autoref{fig:examplesweep} shows the effect of a sweep of the
maximum cell charge current on six key variables.
As can be seen, what may feel like a subsystem-level input can have relatively
dramatic system-wide impacts.
Increasing the allowable charge current reduces charging time which in turn
reduces the number of podbays needed in each portal.
However, increasing the charge current beyond 22 amps does not reduce
turnaround time further because turnaround time is also constrained by
passenger loading/unloading time.
Increasing the charge current also allows the battery to reduce in size quite
substantially, which in turn allows the pod to get lighter.
Meanwhile the optimal cruise velocity and tube pressure also increase.
The sweep's effect on ramp length is particularly interesting.
Initially the optimal ramp length decreases as charge current increases,
because charge current constrains the pod's regenerative braking capability.
Above 16 amps, this is no longer the active constraint for ramp length but a
constraint on maximum instantaneous cell power loss becomes active, which
causes the optimal ramp length to increase again as the optimal cruise velocity
also continues to increase.
Above 20 amps, a constraint on charge voltage becomes active and the optimal
ramp length stabilizes: further increases in charge current do not afford
meaningful cost reduction.
We can identify these constraints becoming active by plotting the sensitivities
to each input as shown in \autoref{fig:sensplot}.
In practice, a sweep like this tells us over which ranges we should care about
the cell charge current limit (from a cost perspective) and why (from a design
perspective).


Because all design relationships are expressed as constraints rather than a
sequence of calculations, sweeps can be performed on variables that would
usually be considered free variables, i.e.  outputs.
For example, a designer can sweep over a range of pod mass values to
understand what the optimal design and performance would be at each point.
This can be helpful for
developing intuition and identifying opportunities where the model is not
adequately capturing true design considerations.

Two-dimensional sweeps (which rely even more on HOPS' speed) can illustrate
relative sensitivities, answering questions
such as ``how much more should you be willing to pay for
track to improve the efficiency of your propulsion system?''

\subsection{Trade studies}
HOPS has been used to perform trade studies between different architectures for
the propulsion, levitation, power electronics, and thermal management systems.
Optimal solutions for two or more discrete architectures can then be compared
to show differences in both high-level cost and subsystem
designs.
For example, the choice of a power electronics technology may significantly
affect the frequency of battery replacement, while the thermal management
system architecture may significantly change the optimal
battery size and pressure management system design.

What HOPS provides is a common platform from which to debate quantitative
merits and drawbacks of each architecture.
HOPS can only provide a ``dry'' cost optimization perspective though, and has
nothing to say on questions of development risk, upgradability, and safety.

\section{Implementation, Transparency, and Trust}

HOPS has been adopted as a central component of the Virgin Hyperloop system
design process.
Given its potential to therefore influence a broad range of significant
decisions, and given that formal system optimization (let alone geometric
programming) is a relatively niche and nascent field in most industries, it has
been important to take extensive measures to develop confidence in
the tool.

This need to build confidence has significantly influenced HOPS' implementation
from a cultural perspective.
The core tenets of this implementation have been transparency, testing, and
distribution of ownership and responsibility.

Finding the right level of involvement has been a challenge, but the most
natural and practical balance has been to have design engineers, manufacturing
engineers, engineering analysts, and business analysts provide models for their
respective domains in whatever format feels most natural to them (usually an
Excel spreadsheet or a MATLAB script) which is then transposed to a
geometric programming formulation and HOPS syntax by a HOPS developer.

\subsection{Documentation}
A key piece of transparency comes from having comprehensive and
up-to-date model documentation.
Given the size of the model and the frequency with which changes are made,
manually updating such documentation would be impractical.
Instead, a custom ``autodocumentation'' tool parses all model code to generate a \LaTeX
document that includes all of the constraints and numerical inputs.
This allows stakeholders to review and check the model without needing to
navigate the code repository or be familiar with Python and GPkit syntax.
HOPS' syntax has been designed to ensure
uniformity and thus enable full documentation coverage with as little
maintenance and customization as possible.
This uniformity has the added benefit of making the code easier to read and
navigate.
Because the documentation is automatically generated with every model change, it is
guaranteed to be up-to-date.

\subsection{Testing}
One challenge of maintaining a large, monolithic, all-at-once optimization tool
is the implementation of effective model testing.
Because of the many inter-model coupling variables, unit testing is difficult
to implement and would itself require a lot of maintenance.
In lieu of unit testing, a suite of full-model test cases are run with each
model change to ensure that all valid model configurations remain functional.

A ``diff'' comparing each new solution to its predecessor is generated
and saved with the test solutions.
These diffs are then manually inspected to
ensure that the effect on each test case is accurate and understood.
Solution visualizations are also used extensively to identify
unexpected or erroneous results.

Finally, HOPS models are cross-validated against other in-house design tools.
For example, system-level pod performance results have been compared with the
outputs of a pod simulation tool and sub-system solution values have been
compared against the original models from which they were generated.

\subsection{Interactivity}
An interactive browser-based tool called iHOPS allows users
across the company to create and solve cases and review solutions without
needing to install HOPS (and all its dependencies) or be familiar with Python
and GPkit.
iHOPS empowers the engineering and business teams to perform
their own point optimizations, sensitivity studies (``sweeps''), or comparisons (``diffs'').
This helps build confidence in HOPS, while also helping to identify any surprising model
behavior or inputs worthy of further investigation.
iHOPS also provides access to key ``reference'' solutions and therefore
serves as a central repository for design values and assumptions.

\subsection{Check-ins}
We encourage widespread buy-in to HOPS results by holding
regular (e.g. monthly) ``check-in'' meetings with all stakeholders to
review models, inputs, and solution values, and to discuss potential
improvements and areas of concern.
These check-ins foster a sense of shared ownership of, and responsibility for,
the development of HOPS and the quality of its solutions.

HOPS is imperfect in many ways.
Despite rigorous efforts to quality-control the models, the possibility of
errors cannot be ruled out.
There are also plenty of high-level inputs that can be described as best
guesses or judgement calls.
Despite this, HOPS serves an important purpose as a unified repository
of information, a tool that aligns members of every engineering team to a
single design point, regardless of whether it is truly optimal\cite{gpkitdis}.
This is only possible because HOPS describes the full hyperloop system:
every subsystem designer's interests (and concerns) are represented.
The purpose of check-ins is to make clear to stakeholders that they have
the power to change that representation at any time.

\section{Conclusion}
The design of a hyperloop system presents an interesting and highly-coupled
optimization problem due to the often-competing objectives of minimizing CapEx,
OpEx, and travel time, and the many recursive design relationships inherent to
the system architecture.
Combining this with the fact it is a clean-sheet design problem makes hyperloop
a unique opportunity to impactfully apply rigorous MDO techniques.

A system optimization tool, HOPS, has been developed that is capable of
providing a fast and disciplined way of solving such an optimization problem, by
formulating it as a sequence of geometric programs that can be solved using
commercially available software.
HOPS minimizes total cost per passenger-km and models everything from the
diameter of the tube down to the current in the motor coils.
The default case has 4512 free variables and solves in 4 seconds on a laptop.

HOPS has played a central role in the Virgin Hyperloop system design process,
and has been to used to optimize the design, set system requirements, perform
sensitivity analysis, and inform trade studies.

\bibliographystyle{aiaa}
\bibliography{hyperloop_system_optimization}

\end{document}

%% file: figs/route_topology.tex
\def\L{10}
\def\r{0.7}
        \begin{tikzpicture}
                \draw[ultra thick] (0,0) node[circle,fill,label=left:$O$](O){} --
                                   (\L,0) node[circle,fill,label=right:$D$](D){};
                \draw[thick] (\L/4,-\r) node[circle,fill,label=below:$A$](A){}
                                        arc (0:90:\r);
                \draw[thick] (\L/4,-\r) arc (180:90:\r);
                \draw[thick] (\L*2/4,\r) node[circle,fill,label=above:$B$](B){}
                                         arc (360:270:\r);
                \draw[thick] (\L*2/4,\r) arc (180:270:\r);
                \draw[thick] (\L*3/4,-\r) node[circle,fill,label=below:$C$](C){}
                                        arc (0:90:\r);
                \draw[thick] (\L*3/4,-\r) arc (180:90:\r);
        \end{tikzpicture}

%% file: figs/velocity_profile.tex
\begin{tikzpicture}
		\draw[thick,->] (0,0)  -- (9,0) node[anchor=north east] {Time};
		\draw[thick,->] (0,0)  -- (0,3) node[anchor=north east] {Velocity};
		\draw         (0.0,0.0)  -- (0.2,0.5);
		\draw[dotted] (0.0,0.5) node[anchor=east] {$v_1$}  -- (0.2,0.5);
		\draw[dotted] (0.2,0.0) node[anchor=north]{$t_1$}  -- (0.2,0.5);
		\draw         (0.2,0.5)  		           -- (0.5,1.0);
		\draw[dotted] (0.0,1.0) node[anchor=east] {$v_{i}$}  -- (0.5,1.0);
		\draw[dotted] (0.5,0.0) node[anchor=north]{$t_{i}$}  -- (0.5,1.0);
		\draw         (0.5,1.0)  			   -- (1.0,1.5);
		\draw[dotted] (0.0,1.5) node[anchor=east] {$v_{i+1}$}  -- (1.0,1.5);
		\draw[dotted] (1.0,0.0) node[anchor=north]{$t_{i+1}$}  -- (1.0,1.5);
		\draw         (1.0,1.5)  			   -- (2.0,2.0);
		\draw[dotted] (1.2,1.6) -- node[below] {$a_n$}	      (1.8,1.6);
		\draw[dotted] (1.8,1.6)  			   -- (1.8,1.9);
		\draw[dotted] (0.0,2.0) node[anchor=east]{$v_\text{c}=v_n$}--(2.0,2.0);
		\draw[dotted] (2.0,0.0) node[anchor=north]{$t_n$}  -- (2.0,2.0);
		\draw         (2.0,2.0)  			   -- (6.0,2.0);
		\draw         (6.0,2.0)  			   -- (7.0,1.5);
		\draw[dotted] (6.2,1.6) -- node[below] {$d_1$}	      (6.8,1.6);
		\draw[dotted] (6.2,1.6)  			   -- (6.2,1.9);
		\draw         (7.0,1.5)  		           -- (7.5,1.0);
		\draw         (7.5,1.0)  			   -- (7.8,0.5);
		\draw         (7.8,0.5)  -- (8.0,0.0);
		\draw [decorate,decoration={brace,amplitude=10pt}]
		      (0,2.1) -- (2.0,2.1) node [black,midway,yshift=15pt] {\footnotesize Launch};
		\draw [decorate,decoration={brace,amplitude=10pt}]
		      (2,2.1) -- (6.0,2.1) node [black,midway,yshift=15pt] {\footnotesize Cruise};
		\draw [decorate,decoration={brace,amplitude=10pt}]
		      (6,2.1) -- (8.0,2.1) node [black,midway,yshift=15pt] {\footnotesize Braking};
\end{tikzpicture}

%% file: figs/velocity_profile_slowdowns.tex
\begin{tikzpicture}
		\draw[thick,->] (0,0)  -- (9,0) node[anchor=north east] {Time};
		\draw[thick,->] (0,0)  -- (0,3) node[anchor=north east] {Velocity};
		\draw         (0.7*0.0,0.0)  -- (0.7*0.1,0.4); 
		\draw         (0.7*0.1,0.4)  -- (0.7*0.2,0.8); 
		\draw         (0.7*0.2,0.8)  -- (0.7*0.35,1.2); 
		\draw         (0.7*0.35,1.2)  -- (0.7*0.55,1.6); 
		\draw         (0.7*0.55,1.6)  -- (0.7*0.8,2.0); 
		\draw         (0.7*0.8,2.0)  -- (0.7*3.2,2.0); 
		\draw         (0.7*3.2,2.0)  -- (0.7*3.45,1.6); 
		\draw         (0.7*3.45,1.6)  -- (0.7*3.65,1.2); 
		\draw         (0.7*3.65,1.2)  -- (0.7*3.8,0.8); 
		\draw         (0.7*3.8,0.8)  -- (0.7*3.9,0.4); 
		\draw         (0.7*3.9,0.4)  -- (0.7*4.0,0.0); 
		\draw         (0.7*4+0.7*0.0,0.0)  -- (0.7*4+0.7*0.1,0.4); 
		\draw         (0.7*4+0.7*0.1,0.4)  -- (0.7*4+0.7*0.2,0.8); 
		\draw         (0.7*4+0.7*0.2,0.8)  -- (0.7*4+0.7*0.35,1.2); 
		\draw         (0.7*4+0.7*0.35,1.2)  -- (0.7*4+0.7*0.55,1.6); 
		\draw         (0.7*4+0.7*0.55,1.6)  -- (0.7*4+0.7*0.8,2.0); 
		\draw         (0.7*4+0.7*0.8,2.0)  -- (0.7*4+0.7*3.2,2.0); 
		\draw         (0.7*4+0.7*3.2,2.0)  -- (0.7*4+0.7*3.45,1.6); 
		\draw         (0.7*4+0.7*3.45,1.6)  -- (0.7*4+0.7*3.65,1.2); 
		\draw         (0.7*4+0.7*3.65,1.2)  -- (0.7*4+0.7*3.8,0.8); 
		\draw         (0.7*4+0.7*3.8,0.8)  -- (0.7*4+0.7*3.9,0.4); 
		\draw         (0.7*4+0.7*3.9,0.4)  -- (0.7*4+0.7*4.0,0.0); 
		\draw         (0.7*8+0.7*0.0,0.0)  -- (0.7*8+0.7*0.1,0.4); 
		\draw         (0.7*8+0.7*0.1,0.4)  -- (0.7*8+0.7*0.2,0.8); 
		\draw         (0.7*8+0.7*0.2,0.8)  -- (0.7*8+0.7*0.35,1.2); 
		\draw         (0.7*8+0.7*0.35,1.2)  -- (0.7*8+0.7*0.55,1.6); 
		\draw         (0.7*8+0.7*0.55,1.6)  -- (0.7*8+0.7*0.8,2.0); 
		\draw         (0.7*8+0.7*0.8,2.0)  -- (0.7*8+0.7*3.2,2.0); 
		\draw         (0.7*8+0.7*3.2,2.0)  -- (0.7*8+0.7*3.45,1.6); 
		\draw         (0.7*8+0.7*3.45,1.6)  -- (0.7*8+0.7*3.65,1.2); 
		\draw         (0.7*8+0.7*3.65,1.2)  -- (0.7*8+0.7*3.8,0.8); 
		\draw         (0.7*8+0.7*3.8,0.8)  -- (0.7*8+0.7*3.9,0.4); 
		\draw         (0.7*8+0.7*3.9,0.4)  -- (0.7*8+0.7*4.0,0.0); 
\end{tikzpicture}

%% file: figs/battery_soc.tex
\begin{tikzpicture}
		\def\boly{5}
		\def\eolx{2}
		\def\eoly{2.2}
		\def\eocx{7}
		\def\eocy{1.6}
		\def\eorx{9}
		\def\eory{4}
		\def\eolix{8}
		\def\eoliy{1.2}
		\def\maxnomy{0.8}
		\def\maxoffnomy{0.4}
		\draw[thick] (0,0) node[anchor=east, scale=1.2]{0\%}  -- (\eorx,0) node[anchor=west, scale=1.2]{100\%} ;
		\draw[thick,->] (0,0)  -- (0,\boly*1.2) node[anchor=south east,scale=1.5] {SoC};
		\draw[thick,<-] (\eorx,0)  -- (\eorx,\boly*1.2) node[anchor=south west,scale=1.5] {DoD};
		\draw[dotted] (0,\boly/7*8) node[anchor=east, scale=1.2]{100\%}  -- (\eorx,\boly/7*8) node[anchor=west, scale=1.2]{0\%} ;
		\draw (0.0,\boly) -- (\eolx,\eoly);
		\draw[dotted] (\eolx,0.0) -- (\eolx,\eoly);
		\draw[<->] (0,-0.4) -- node[fill=white, scale=1.1]{Launch} (\eolx,-0.4);
		\draw         (\eolx,\eoly)  			   -- (\eocx,\eocy);
		\draw[<->] (\eolx,-0.4) -- node[fill=white, scale=1.1]{Cruise} (\eocx,-0.4);
		\draw         (\eocx,\eocy)  			   -- (\eorx,\eory);
		\draw[dotted] (\eocx,0.0) -- (\eocx,\eocy);
		\draw[dotted] (\eocx,\eocy)   -- (\eorx,\eocy)node[anchor=west, scale=1.2]{$DoD_{\text{nom}}$};
		\draw[dotted] (\eorx,\eory) --  (\eorx,\eory) node[anchor=west, scale=1.2]{$DoD_{\text{final}}$};
		\draw[dashed]        (\eocx,\eocy)  -- (\eolix,\eoliy);
		\draw[dotted] (\eolix,0.0) -- (\eolix,\eoliy);
		\draw[<->,dashed] (\eocx,-0.9) -- node[fill=white, scale=1.1, yshift=-10pt]{Limp} (\eolix,-0.9);
		\draw[<->] (\eocx,-0.4) -- node[fill=white, scale=1.1]{Braking} (\eorx,-0.4);
		\draw[dotted] (\eolix,\eoliy)  -- (\eorx,\eoliy)
					 node[anchor=west, scale=1.2]{$DoD_{\text{offnom}}$} ;
		\draw[<->] (-0.5,\boly) -- node[rotate=90, fill=white, scale=1.2]{$DoD_{\text{nom, max}}$} (-0.5,\maxnomy);
		\draw[dotted] (0,\maxnomy)   -- (\eorx,\maxnomy);
		\draw[<->] (-1.0,\boly) -- node[rotate=90, fill=white, scale=1.2]{$DoD_{\text{offnom, max}}$} (-1.0,\maxoffnomy);
		\draw[dotted] (0,\maxoffnomy)   -- (\eorx,\maxoffnomy);
\end{tikzpicture}